\pdfoutput=1

\newcommand{\caproman}[1]{\uppercase\expandafter{\romannumeral#1}}

\newcommand{\BGmail}{bernardgottschalk\,@\,gmail.com}

\newcommand{\sect}[1]{Sec.\,\ref{sec:#1}}
\newcommand{\eqn}[1]{Eq.\,(\ref{eqn:#1})}

\newcommand{\fig}[1]{Fig.\,\ref{fig:#1}}

\documentclass[12pt]{article}

\title{\bf Measuring the length of stretched wires\\with an avalanche oscillator}
\author{B. Gottschalk\thanks{Harvard University Laboratory for Particle Physics and Cosmology (LPPC), 18 Hammond St., Cambridge, MA 02138, USA, \BGmail}}
\date{\today}

\usepackage{amsmath}
\usepackage{latexsym}
\usepackage{graphicx}
\usepackage{rotating}
\usepackage{lineno}
\usepackage{setspace}
\usepackage[linktocpage=true,pdftex,hyperfigures=true]{hyperref}

\begin{document}

\maketitle



\begin{abstract}
\noindent We measure the length of stretched wire pairs by using them as the delay line pulse shaping element in an avalanche oscillator. The circuitry and method are extremely simple, and insensitive to oscillator supply voltage, to the particular transistor (2N3904) used and to wire tension. Rudimentary tests with simulated broken wires show that these can be detected easily. It remains to be seen whether the technique will work at scale with realistic wire planes, but it should be at least as good as reflecting short pulses.
\end{abstract}

\tableofcontents

\section{Introduction}\label{sec:Introduction}
Modern particle detectors employ planes with thousands of wires. Confirming their integrity includes testing for broken wires or failed solder joints. It has been proposed to measure the delay time, down and back, of very short pulses launched at one end. While that will almost certainly work, we investigate a simpler scheme here. We use a stretched wire pair as the pulse shaping network in an avalanche oscillator, chosen for its simplicity and the very short risetimes it produces. Then, only the width of the resulting pulse need be measured.

In this proof-of-principle we describe the equipment (oscillator, oscilloscope and high voltage (HV) supply) and tests we have done to measure the sensitivity of pulse width to HV, to the particular transistor used and to wire length. Finally, we perform some very rudimentary tests with mock broken wires. These results show the method is practical and invite studies of larger scale and more realistic configurations.

An Appendix covers the basics of delay lines used as pulse shaping elements.

 \section{Equipment}\label{sec:Equipment}
\subsection{Oscillator}
\begin{figure}[p]
\centering\includegraphics[height=4in]{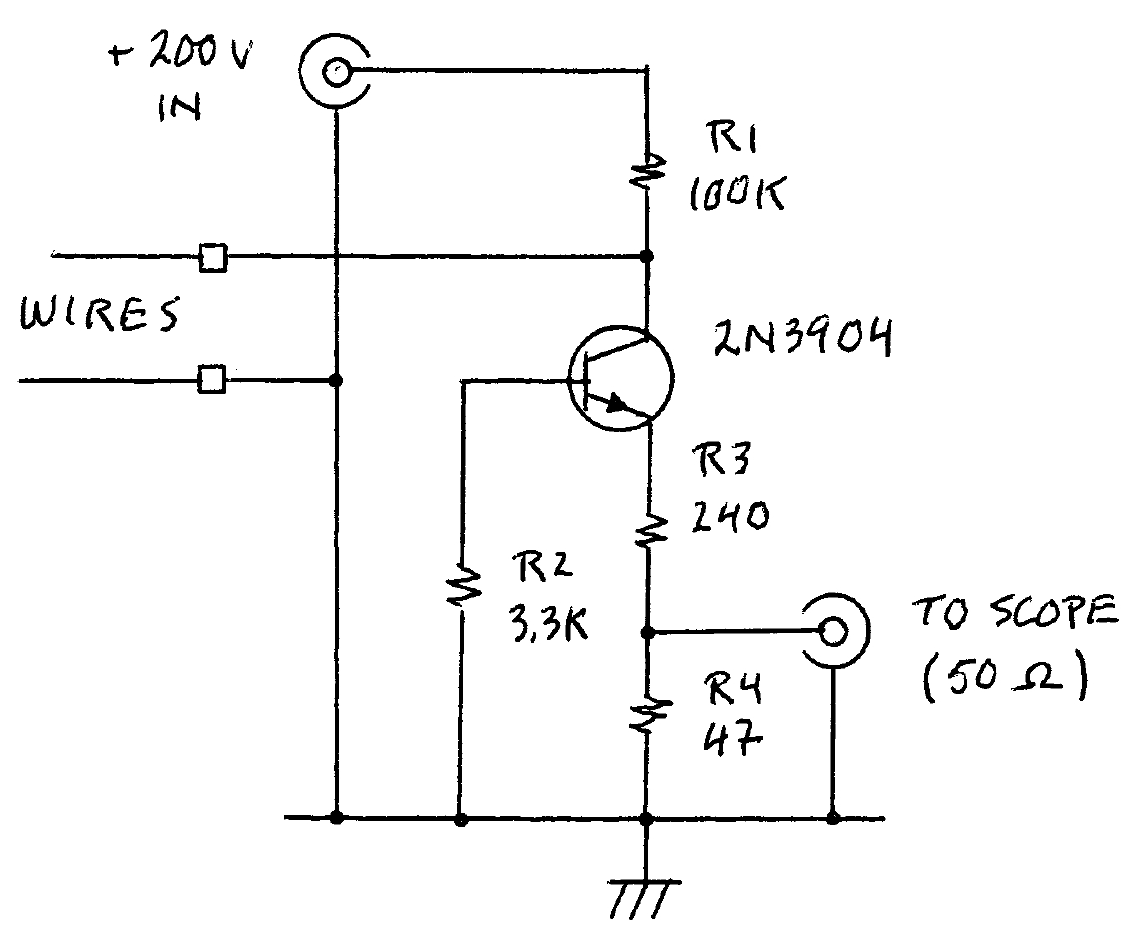}
\caption{Schematic diagram. Resistors are 1/4w 5\%.}\label{fig:schematic}
\end{figure}

\begin{figure}[p]
\centering\includegraphics[height=3.8in]{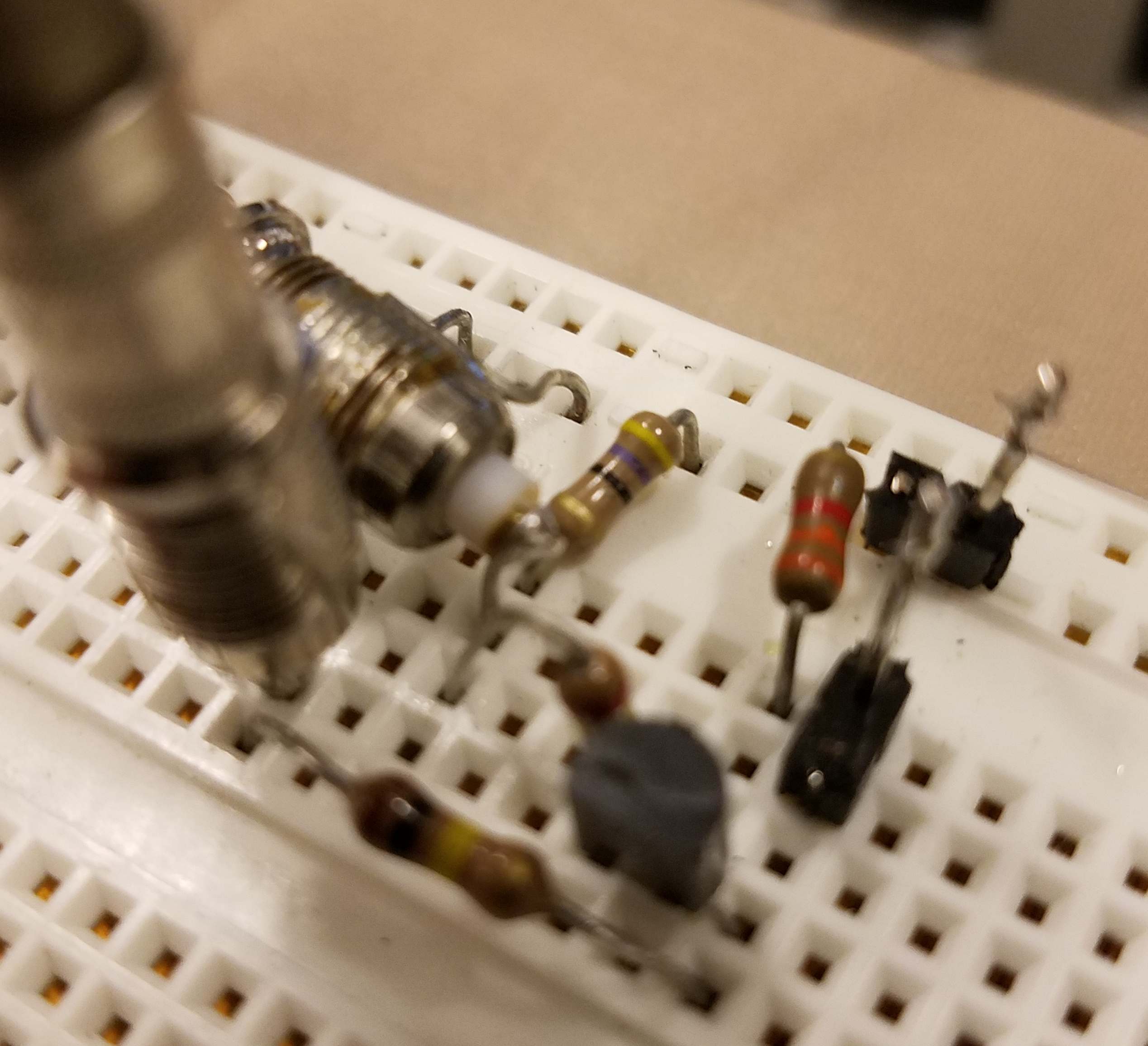}
\caption{Circuit as built: note short leads. Connectors are KINGS-LEMO\,1074-1 with leads soldered to center pin and shell. Scope is connected via Lemo-to-BNC adapter and a 20\,cm ($\approx1$\,ns) RG58C/U cable.}\label{fig:construction61}
\end{figure}

\fig{schematic} is a schematic diagram of the oscillator and \fig{construction61} shows its construction using legacy breadboard technique. The circuit follows Holme \cite{Holme2003}, adjusting for the much greater characteristic impedance of the parallel-wire line.

The avalanche oscillator works as follows. Assume the collector is initially low. The delay line charges exponentially towards 200\,V until it reaches $V_\mathrm{\,breakdown}$  when avalanche multiplication of carriers occurs in the depletion layer. That short-circuits the collector to the emitter and connects the high side of the delay line (initially at $V_\mathrm{\,breakdown}$) to the 240\,$\Omega$ resistor. Discharging the line takes twice the one-way transit time and generates $0.5\,V_\mathrm{\,breakdown}$ at the load (see Appendix \ref{sec:Basics}). That removes the energy source for the avalanche, the transistor returns to its high impedance state with the collector still low, and the cycle repeats.

The 240\,$\Omega$ resistor was chosen by trial and error to minimize reflections, and the output flat-top was found to be $\approx2.5$\,V. That bespeaks
\begin{equation}
V_\mathrm{\,breakdown}=2\times\frac{20+240+25}{25}\times2.5=57\mathrm{\,V}
\end{equation}
which is reasonable since $V_\mathrm{CE}$ (max) is 40\,V for the 2N3904. We have assumed the transistor adds about 20\,$\Omega$ following Holme \cite{Holme2003}. 
\footnote{~It is difficult to measure the sawtooth at the collector directly. The mere application of a probe changes the output pulse drastically indicating the observed waveform is invalid.}

Holme operates at a much lower HV\,$\approx130$\,V and breakdown occurs much later in the exponential rise. In our circuit under those conditions the output voltage was greater, but less stable from pulse to pulse. At 200\,V breakdown occurs on the linear part of the exponential rise and pulse width is nearly independent of applied voltage (see \sect{HVtests}).

 \subsection{HV Supply}
This was a TEXIO PA600-0.1B regulated DC power supply. We observed some baseline ringing after the pulse when we used clip leads for the connection. Coaxial RG-174B/U cable works better, and no additional bypass seems to be necessary. The HV was set to 200\,V except as noted.
 
\subsection{Oscilloscope}
This was a Lecroy Wave Runner 204MXi 2\,GHz. The Operator's Manual gives the calculated risetime as 225\,ps, well below anything we observed. We set the scope up to display pulse width (evidently, full width at half maximum), one of the built-in measurements.

\subsection{Setup and Procedure}
\begin{figure}[p]
\centering\includegraphics[height=4in]{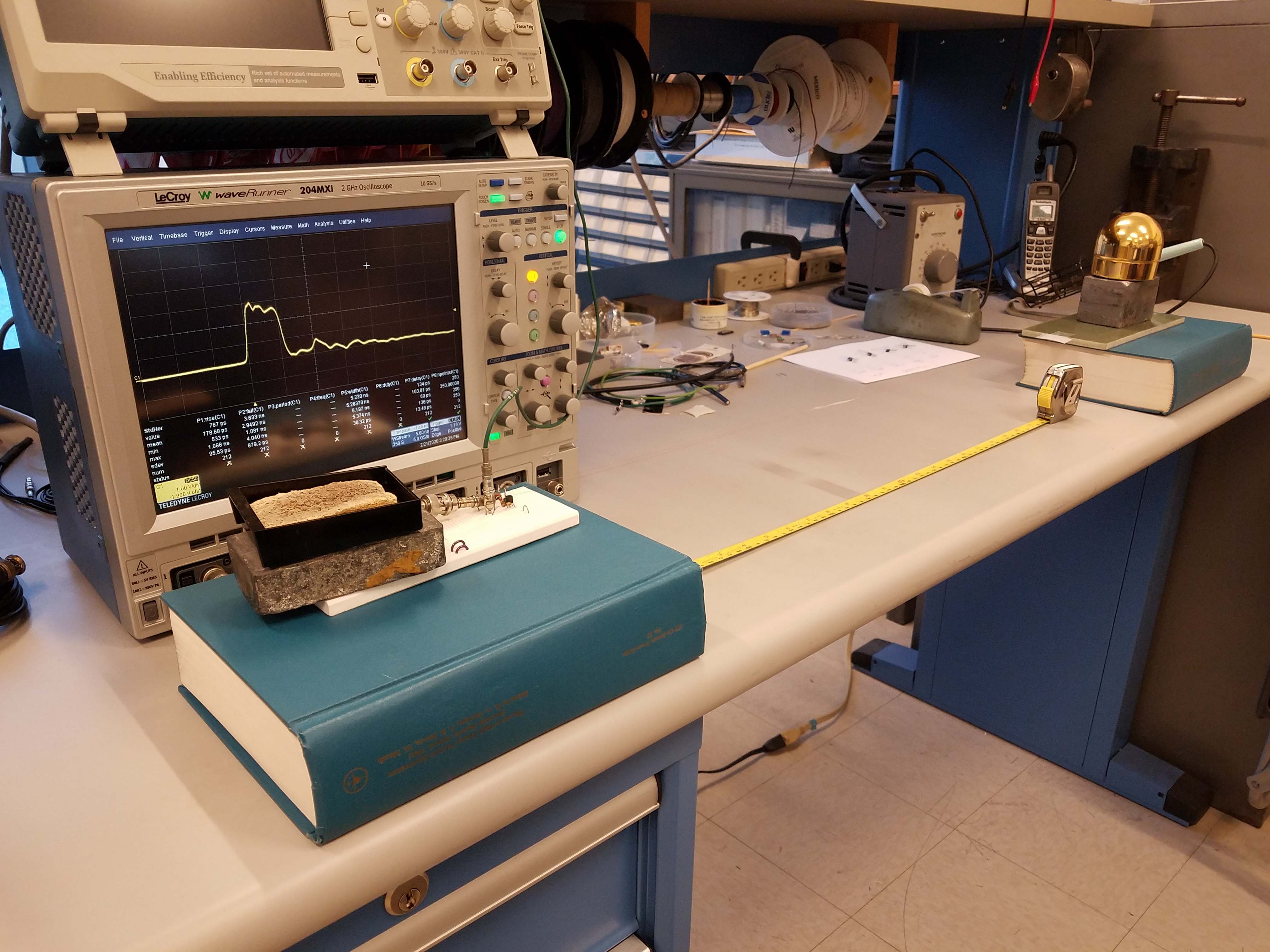}
\caption{Typical test setup.}\label{fig:brokenSetup61}
\end{figure}

A typical setup is shown in \fig{brokenSetup61}. The wires are soldered to pins at the circuit end and trapped between Scotch Magic$^\mathrm{\scriptscriptstyle{TM}}$ tape at the far end. Pulse width is extremely insensitive to tension and for that matter, wire spacing. The wires can be slack as long as they do not touch.

For the illustrations below we used scope ACQ mode which displays the average of a number of waveform acquisitions. For a typical measurement we cleared sweeps (which also clears results), pressed NORMAL to start measuring, pressed STOP when approximately 200 shots had been analyzed (a second or two), and wrote down the results.

\section{Tests and Results}\label{sec:Tests}
Errors are $1\sigma$ except as noted.

\subsection{Width vs. HV}\label{sec:HVtests}
We investigated the sensitivity of pulse width to HV varying it in the sequence 200, 180, 160, 220, 200\,V. The wire length was 70.33\,cm and the nominal output pulse observed was 2.5\,V except at 160\,V where it was substantially larger. The average of the remaining four readings was $5.088\pm0.028$\,ns ($\pm0.6$\%). Even at 160\,V the width was only 2\% smaller. Thus, pulse width is rather insensitive to HV around 200\,V.

\subsection{Width vs. Transistor}
We investigated the sensitivity of pulse width to the particular 2N3904 used, picking 5 transistors from a bin at random.\footnote{~The 2N3904, introduced by Motorola 55 years ago, is still widely used and still easily available.} Conditions were the same as the preceding test. Four units yielded 2.5\,V output, the fifth 3.1\,V. The average pulse width over all was $5.036\pm0.052$\,ns ($\pm1.0\%$). Thus, pulse width is rather insensitive to the particular transistor used. A single transistor `\#1' was used for the remaining tests.

\subsection{Width vs. Length}
\begin{figure}[p]
\centering\includegraphics[height=4in]{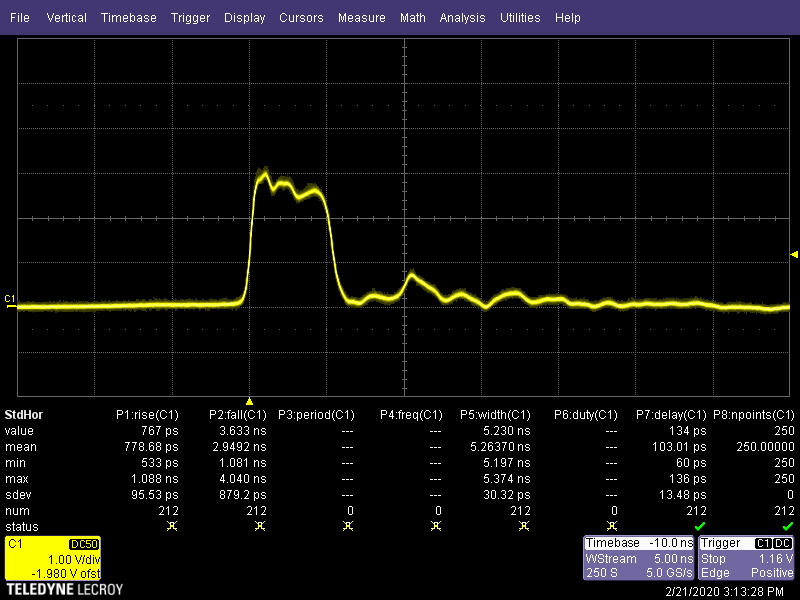}
\caption{Screenshot, 73\,cm wire length.}\label{fig:73cmSS}
\end{figure}

\begin{figure}[p]
\centering\includegraphics[height=4in]{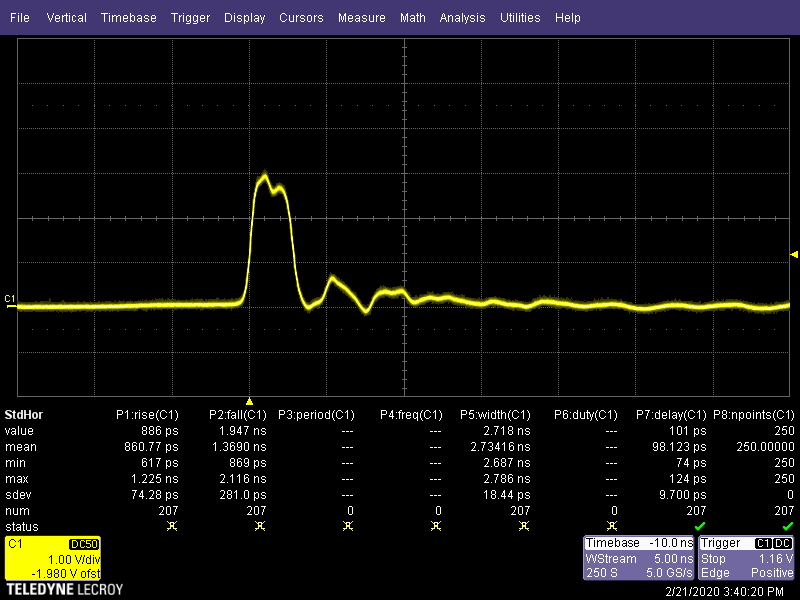}
\caption{Screenshot, 34\,cm wire length.}\label{fig:34cmSS}
\end{figure}

\begin{figure}[p]
\centering\includegraphics[height=4in]{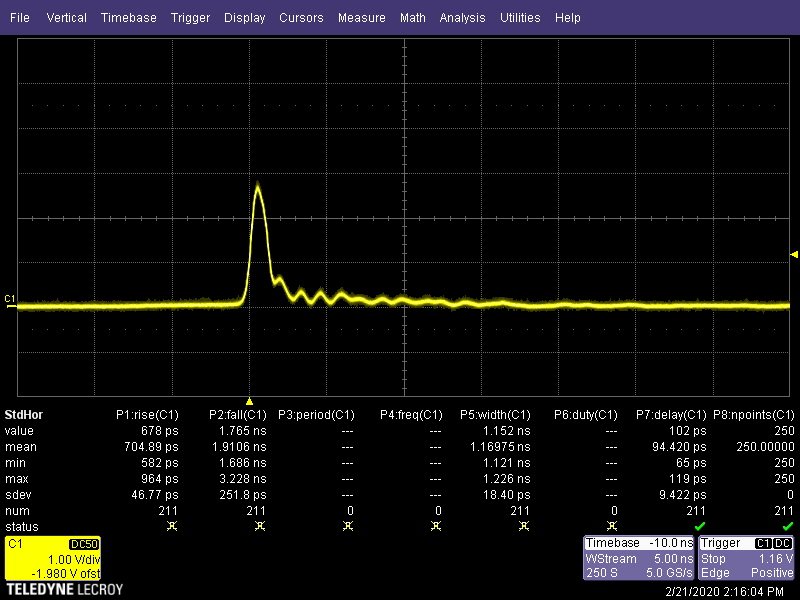}
\caption{Screenshot, 10\,cm wire length. This pulse is 1.17\,ns wide.}\label{fig:10cmSS}
\end{figure}

We cut the wires back, mostly in $\approx12$\,cm steps. Representative screenshots at 73, 34 and 10\,cm length are shown in Figs. \ref{fig:73cmSS}, \ref{fig:34cmSS} and \ref{fig:10cmSS} respectively. At each length, at least 3 measurements were made, each taking a few seconds and consisting of $\approx200$ analyses by the scope. The measurements were then averaged.

\begin{figure}[p]
\centering\includegraphics[height=4in]{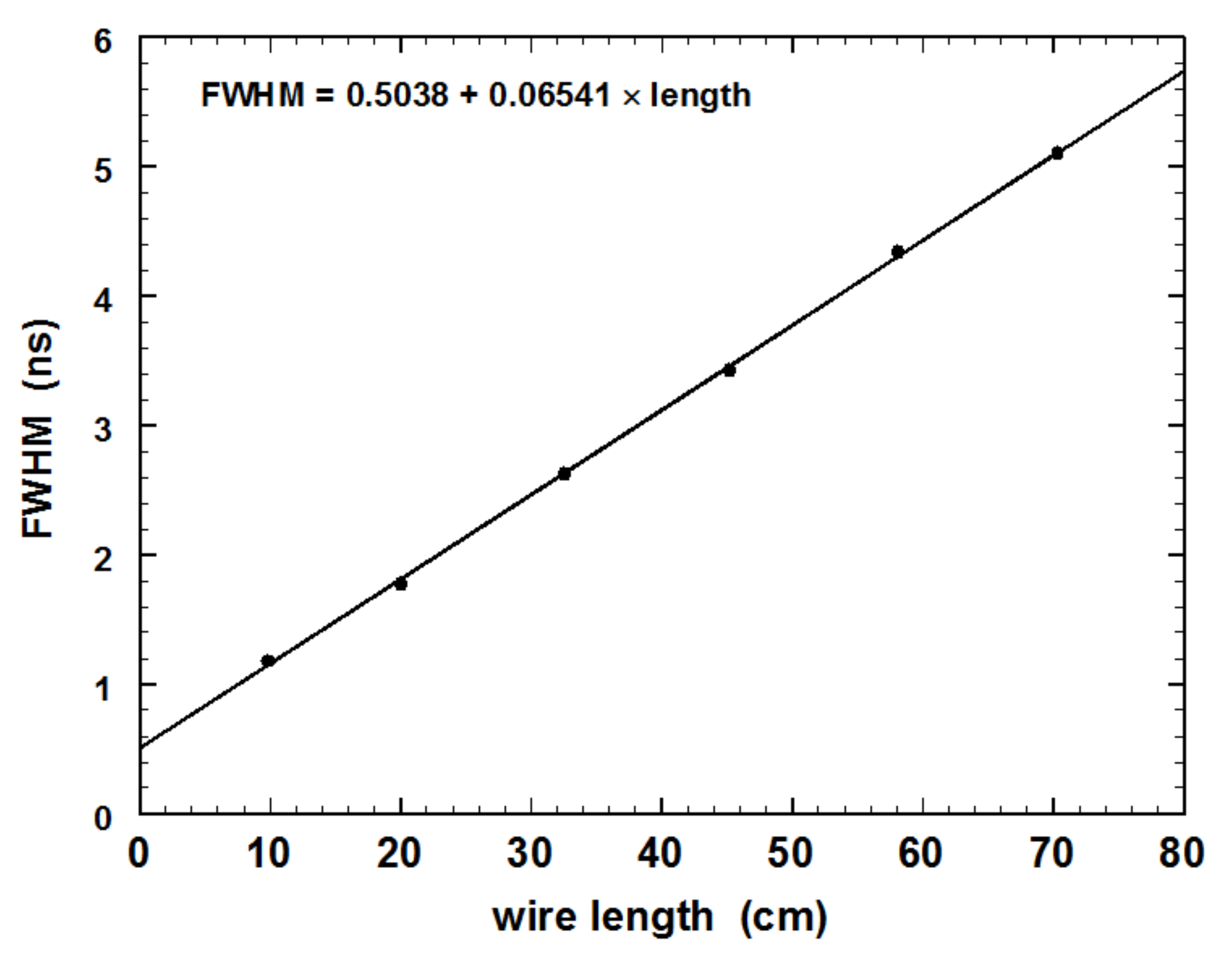}
\caption{Length measurements with fitted line.}\label{fig:nsVcm}
\end{figure}

\fig{nsVcm} shows the results with a straight line fit. The standard ($1\sigma$) fit error per point, from residuals, is 
\begin{equation}\label{eqn:S}
S\equiv\sqrt{\sum{(y_i-y(x_i))^2/(N-2)}}=0.0332\mathrm{\,ns}
\end{equation}
The slope, $0.06541\pm0.00065$\,ns/cm $(\pm1.0\%)$, equals $2\tau/l$ (see Appendix \ref{sec:Basics}) where $\tau$ is the one-way transit time and $l$ is the wire length. Therefore the measured propagation speed is
\begin{equation}\nonumber
U=\frac{l}{\tau}=\frac{2}{\mathrm{slope}}=30.57\,\,\frac{\mathrm{cm}}{\mathrm{ns,}}
\end{equation}
or $1.019\,c$ ($c\equiv$ speed of light $=30$\,cm/ns). Thus, more or less within experimental error, we confirm the prediction for an air dielectric line \cite{Terman1943}.

Converting $\sigma_\mathrm{slope}$ to an equivalent error in length we find $\sigma_\mathrm{length}=0.51$\,cm. Changes in length of a few cm should be easily detected.

\subsection{Broken Wires}
Finally, we investigated broken wires directly. The setup (\fig{brokenSetup61}) was shown earlier. The wires were trapped in Scotch Magic$^\mathrm{\scriptscriptstyle{TM}}$ tape at the far end, and also near the middle (simulating the `combs' proposed in actual wire arrays). Since this test is unrealistic at best (see \sect{Discussion}) we only ran four cases, averaging four measurements of $\approx200$ scope samples for each case.  

{\bf Case 1} was the unbroken 73.3\,cm delay line. In {\bf case 2} the hot wire was cut at the far end and left to dangle on the bench. In {\bf case 3} it was cut off at the `comb'. Finally in {\bf case 4}, the remaining cold (ground) wire was also cut leaving an intact line about half the original length. The results, with $1\sigma$ errors derived from the scatter in four measurements, are
\begin{description}
\item{\bf case 1:} $5.2667\pm0.0028$\,ns ($\pm0.05\%$)
\item{\bf case 2:} $4.7289\pm0.059$\,ns ($\pm1.2\%$)
\item{\bf case 3:} $2.6998\pm0.0098$\,ns ($\pm0.4\%$)
\item{\bf case 4:} $2.7323\pm0.028$\,ns ($\pm1.0\%$)
\end{description}
The rather small decrease in width (0.54\,ns) when the cut wire dangles fits in with our early observation that pulse shape and width are nearly impervious to the distance between wires. That in turn is probably because the characteristic impedance \cite{Terman1943} 
\begin{equation}
Z_0=276\,\mathrm{log}_{10}\left(\frac{s}{r}\right)
\end{equation}
($s\equiv$ wire spacing, $r\equiv$ wire radius) depends only logarithmically on wire spacing. Even so, the change in width (0.54\,ns) is $16\times$ greater than the rms error from \eqn{S}. A dangling half-wire is easily detected in our simple setup.

\section{Discussion and Summary}\label{sec:Discussion}
It will not have escaped the reader that our proof-of-principle is highly idealized. A real wire plane is dense with wires and a broken wire will certainly touch many others. We have not looked at the effect of neighboring wires, traces and connectors at the wire ends, other components that might be installed, or how to use our technique at scale.

However, all those complications also beset the more obvious technique of looking at the reflections of very short pulses. The simpler method described here should work at least as well.

\section{Acknowledgements}
We thank Nathan Felt for lending us the oscilloscope, and Harvard University, the Physics Department, and the Laboratory for Particle Physics and Cosmology (LPPC) for ongoing support.

\appendix
\section{Delay Line Basics}\label{sec:Basics}
The general theory of transmission lines  is complicated. See for instance Terman \cite{Terman1943} or Moskowitz and Racker \cite{Moskowitz1951}, sources of the formulas below. Fortunately we need only a tiny subset: a parallel-wire air line, nearly lossless, used as a pulse shaping element. Consider a line of physical length $l$, electrical length (one-way delay time) $\tau$ and characteristic impedance $Z_0$. First let us sketch qualitatively how it works.

\begin{figure}[p]
\centering\includegraphics[height=4in]{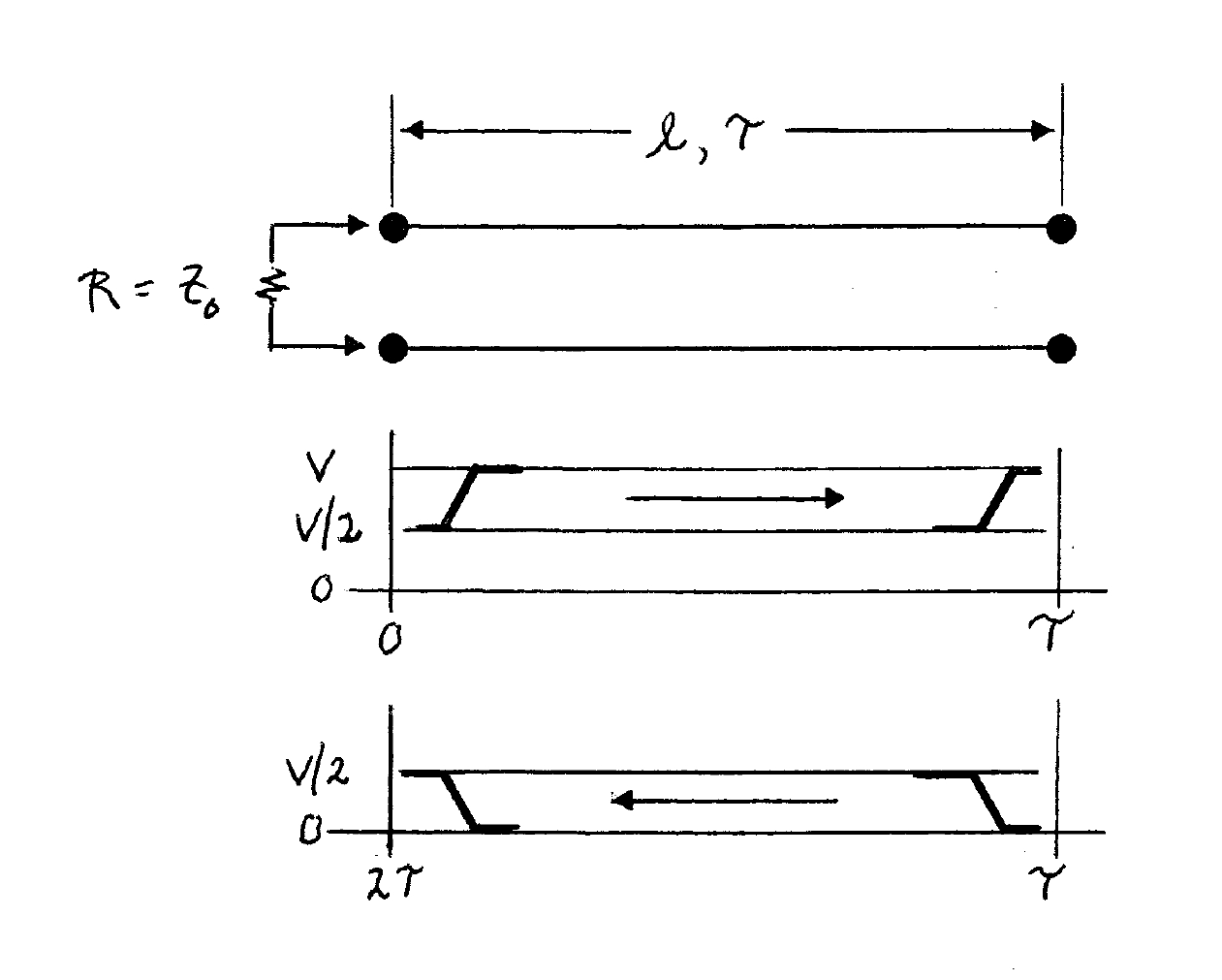}
\caption{Delay line as a pulse shaping element.}\label{fig:waveDiagram}
\end{figure}

Suppose the upper wire in \fig{waveDiagram} has somehow been charged to a voltage $V$ and that, some time later at time $t=0$, the terminating resistor of value $R=Z_0$ is attached. One might guess the outcome across $R$ to be a voltage pulse $V$ of duration $\tau$, but one would be wrong. Since the line acts as a voltage source of impedance $Z_0$, the voltage at the left end immediately drops to $V/2$, but that is not true further right. Rather, a voltage edge from $V$ to $V/2$ travels to the right at a speed $U$ reaching the right end at $t=\tau$ and only then is the voltage throughout the line reduced to $V/2$. Since the right end is open, the edge is reflected in phase and travels back reducing $V/2$ to $0$ as it goes. At $t=2\tau$ it reaches the left end having discharged the line. The pulse across $R$ is therefore $0.5\,V$ with duration $2\tau$. 

Now let us show that, if and only if the terminating resistor $R$ equals $Z_0$, all the energy in the line is dumped into $R$ so that there are no reflections. Let $R$ have any value. $Z_0$ is given by
\begin{equation}\label{eqn:Z0}
Z_0=\sqrt{\frac{L}{C}}
\end{equation}
$L, C$ are {\em total} inductance and mutual capacitance.\footnote{~NB. our notation here differs from the references.} The speed of a wave along the line is
\begin{equation}\label{eqn:U}
U=\frac{l}{\sqrt{LC}}
\end{equation}
The one-way delay, eliminating $L$ from these equations, is therefore
\begin{equation}\label{eqn:tau}
\tau\equiv\frac{l}{U}=Z_0\,C
\end{equation}
The energy dissipated in $R$ is (current$\times$voltage$\times$time) where the voltage in the general case is $VR/(R+Z_0)$ and the time is $2\tau$ (not $\tau$) as shown previously. That leads to
\begin{equation}\label{eqn:energy1}
\mathrm{energy}=2\,C\,V^2\frac{R\,Z_0}{(R+Z_0)^2}
\end{equation}
Finding, in the usual way, the maximum of the fraction with respect to $R$ yields $R=Z_0$ for which
\begin{equation}\label{eqn:energy2}
\mathrm{energy}=\frac{1}{2}\,C\,V^2
\end{equation}
which is indeed the energy stored in a capacitor $C$ charged to $V$. For any other $R$ there will be pulse reflections damped (but not totally) by $R$ and losses in the line.

\end{document}